\documentclass[journal]{IEEEtran}
\usepackage{amssymb}
\usepackage{amsmath}      % math equation package
\usepackage{cite}         % cite reference package
\usepackage{algorithmicx} % algorithm table??
\usepackage{graphicx}     % quote figure
\usepackage{tabularx, booktabs} % build up table
\usepackage{nomencl}
\makenomenclature
\usepackage{subfig}
\usepackage{caption}
\usepackage{xcolor} % text color
\usepackage{color}
\usepackage{hyperref}
\usepackage{multirow}
\DeclareSubrefFormat{parens}{#1(#2)}
\usepackage{tikz}
\usetikzlibrary{arrows.meta,positioning,shapes.geometric}

\usepackage{booktabs,multirow,array}

\newcolumntype{L}[1]{>{\raggedright\arraybackslash}p{#1}}
\newcolumntype{C}[1]{>{\centering\arraybackslash}p{#1}}
\newcolumntype{R}[1]{>{\raggedleft\arraybackslash}p{#1}}

\begin{document}

\title{PFAgent: A Tractable and Self-Evolving Power-Flow Agent for Interactive Grid Analysis}

\author{Buxin She,~\IEEEmembership{Member,~IEEE,}
Brian Chen,
Luanzheng Guo,~\IEEEmembership{Member,~IEEE,}
and Fangxing Li,~\IEEEmembership{Fellow,~IEEE}

\thanks{B. She is with the Mike Wiegers Department of Electrical and
Computer Engineering, Kansas State University, Manhattan, KS 66506 USA.}
\thanks{B. Chen and L. Guo are with Pacific Northwest National Laboratory, Richland, WA 99354 USA.}
\thanks{F. Li is with the Department of Electrical Engineering and Computer Science,
The University of Tennessee, Knoxville, TN 37996, USA.}
\thanks{Corresponding author: Buxin She (shebuxin@gmail.com).}
        } % Stop of the author part

% ------------------------- headers, abstract, keywords ------------------------
% The paper headers
\markboth{Preprint}%
{She \MakeLowercase{\textit{et al.}}: PFAgent: A Tractable and Self-Evolving Power-Flow Agent for Interactive Grid Analysis}

% make the title area
\maketitle

\begin{abstract}
Power system simulation workflows remain expert-intensive. Engineers must translate study intents into code or API calls, execute analyses, and interpret outputs. 
To automate this workflow, this paper presents \emph{PFAgent}, a tractable and self-evolving power-flow agent for interactive grid analysis.
\emph{PFAgent} integrates four key capabilities:
i) a tractable and interactive architecture for intent parsing, knowledge retrieval, tool execution, and structured reporting;
ii) a self-evolution mechanism combining verification-driven refinement and human-in-the-loop feedback;
iii) an AI-assisted evaluation and debugging loop that leverages conversational context, generated code, and execution errors for iterative fixing;
and iv) an evaluation framework covering task success, convergence validity, numerical consistency, and explanation quality.
Verification on IEEE benchmark systems shows that \emph{PFAgent} can automate case change, analyze voltage violations, perform N--1 contingency analysis, generate plots and concise summaries, and return reproducible results with transparent execution logs.
The proposed framework highlights a shift from conventional simulation tools to interactive, tractable, and self-evolving agents for power system analysis.
\end{abstract}


\IEEEpeerreviewmaketitle

% ========================================================================================
% Section I: Introduction
% ========================================================================================

\section{Introduction}

\IEEEPARstart{P}{ower}-system study workflows are being reshaped by the high penetration of inverter-based resources~\cite{she2023virtual}, more frequent extreme events that stress resilience operation~\cite{panteli2017power}, fast-growing data-center and electrified loads that alter demand profiles~\cite{masanet2020recalibrating}, and tighter planning-to-operation feedback loops~\cite{kroposki2017achieving}.
Each of these pressures forces analysts to run many scenario iterations over changing network states and user-defined models~\cite{she2026hybrid}.
A typical study is procedural as well as numerical. 
An engineer has to select the correct benchmark or uploaded case, modify loads, setpoints, or topology, rerun power flow, inspect voltages and line flows,
capture plots and tables, and save outputs for later comparison.
Much of this effort is not the computation itself, but the surrounding tool specifications, case management, and reproducible reporting.

Large language models (LLMs) offer a natural-language interface to this workflow, but an engineering assistant is fundamentally different from a general-purpose chatbot~\cite{cheng2025leveraging, zhao2026large}.
It has to follow exact simulator APIs, respect the ordering of a power-system workflow, preserve cumulative modifications across turns, and expose enough structure that its numerical claims can be reviewed against executable code~\cite{jin2025gridmind}.
In other words, conversational quality is not sufficient. Execution correctness, transparency, and reproducibility are first-class requirements.
This paper uses the term \emph{power-system agent} to describe systems that convert natural-language study objectives into admissible simulator actions and auditable study outputs under such domain-specific constraints.

Recent literature motivates this framing.
Both the opportunities and the limitations of LLMs are discussed in electric-energy systems, emphasizing that domain grounding, tool use, and safety remain essential for practical deployment~\cite{majumder2024exploring, huang2024large}.
Around these perspectives, task-oriented studies have shown that LLMs can assist with fault diagnosis in power grids~\cite{jing2024fault}, cybersecurity analysis of smart-grid applications~\cite{zaboli2024chatgpt}, and retrieval-augmented carbon-footprint accounting~\cite{wang2024carbon}.
These works establish that LLMs are useful in the power and energy domain, but they do not solve the implementation problem of reliable text-to-simulation.
Even for simple load-flow code generation, general-purpose models hallucinate technical details and silently alter network specifications~\cite{bonadia2023potential}, and introducing
natural-language interfaces raises nontrivial security and reliability concerns that cannot be mitigated by language quality alone~\cite{ruan2024threats}.

More recent work wraps LLMs into tool-using agents to meet engineering workflow requirements.
Modular prompting and tool feedback can substantially improve LLM performance on previously unseen simulation tools, as shown on Daline~\cite{jia2024daline}.
A feedback-driven multi-agent framework in~\cite{jia2025enhancing} integrates retrieval, reasoning, and environmental-acting modules with an error-feedback loop, and reports strong results on Daline and MATPOWER benchmarks.
General LLM-systems building blocks such as retrieval-augmented generation~\cite{lewis2020rag} and reasoning-and-acting frameworks~\cite{yao2023react} underpin these efforts.

Existing works mark clear progress, yet several gaps remain before such agents can serve as practical power system assistants.
First, machine-checkable results are uncommon. Models return text or code, but not verifiable structured outputs.
Second, existing agents are largely static in their prompt templates and tool configurations.
They lack a systematic mechanism that attributes verified failures, updates constraints, and incorporates deployment feedback so that the agent can continuously evolve over successive verification.
Third, when generated code fails at runtime, error recovery is typically a shallow regeneration on the same prompt.
No existing power-system agent provides a built-in AI-assisted error-fixing capability that uses the full conversational context, the exact failing code, and the execution error together with repository-level knowledge to produce a repair.
Table~\ref{tab:related_positioning} summarizes how representative prior works relate to these requirements.

\begin{table*}[!t]
\caption{\emph{PFAgent} relative to representative prior work on LLMs in power systems.}
\label{tab:related_positioning}
\centering
\footnotesize
\setlength{\tabcolsep}{3pt}
\renewcommand{\arraystretch}{1.15}
\begin{tabular}{C{0.08\textwidth} L{0.25\textwidth} C{0.12\textwidth} C{0.17\textwidth} C{0.15\textwidth} C{0.16\textwidth}}
\toprule
\textbf{Ref.} &
\textbf{Problem addressed} &
\textbf{Executes external simulator} &
\textbf{Machine-checkable Output} &
\textbf{Self-evolution from verified failures \& feedback} &
\textbf{Built-in AI-assisted error fixing} \\
\midrule
\cite{majumder2024exploring,huang2024large} &
Literature review on LLM use in energy systems &
No & No & No & No \\
\cite{jing2024fault,zaboli2024chatgpt,wang2024carbon} &
Power system diagnosis, cybersecurity, and knowledge-retrieval workflows &
Not the core task & Task-specific only & No & No \\
\cite{bonadia2023potential} &
Single-turn code generation for OpenDSS power flow studies &
Yes & No & No & No \\
\cite{jia2024daline} &
Simulation improvement on a previously unseen simulator (Daline) &
Yes & Tool-feedback scoring, no user-facing contract & Static tool feedback only & No \\
\cite{jia2025enhancing} &
Cross-tool simulation improvement through retrieval, reasoning, acting, and error feedback &
Yes & Benchmark success metrics &
Error-feedback loop; no failure-driven rule update &
Regeneration via error signal; no repository repair \\
\textbf{\emph{PFAgent}} &
\textbf{Tractable design, self-evolving mechanism, and error recovery} &
\textbf{Yes} & \textbf{Yes, six-aspect scoring criteria} &
\textbf{Yes, updates from failure attribution and user feedback} &
\textbf{Yes, context-aware AI fixing with iterative local validation} \\
\bottomrule
\end{tabular}
\end{table*}

To bridge these gaps, this paper presents \emph{PFAgent}, a tractable
and self-evolving power-flow agent for interactive grid analysis.
\emph{PFAgent} targets text-to-power-flow simulation and is built on
ANDES~\cite{cui2021hybrid}, an open-source Python library for power
system analysis. ANDES has a Python API that includes topology and power-flow objects needed for programmatic case editing, N--1 analysis, and result inspection.
The agent is organized as a modular pipeline behind a user
interface. It parses study intents, retrieves context from the
official ANDES manual and case metadata, generates and executes
validated Python on an isolated environment, and returns a structured
report together with traceable logs and plot results.
On top of this pipeline, a self-evolution mechanism turns failures in verification and deployed sessions into updates of
the agent's prompts and rules.
An AI-assisted fixing loop uses conversation context, the failing code, and the execution error to repair code with retrieval in real time.
The main contributions of this work are as follows.
\begin{itemize}
    \item A modular agentic architecture that supports interactive and tractable text-to-simulation workflows through intent parsing, knowledge retrieval, tool execution, and structured reporting, together with prompt engineering, fine-tuning, and a dedicated user interface.
    \item A self-evolution mechanism that integrates evaluation, verification, automatic updating, and human-in-the-loop judgment for continuous agent improvement.
    \item An AI-assisted evaluation and fix loop that sends
    conversational context, generated code, and execution errors to the debugging pipeline for iterative correction and refinement.
    \item An evaluation method for power-system agent performance, including task success, convergence validity, numerical consistency, and explanation quality.
\end{itemize}

The proposed techniques are implemented in \emph{PFAgent}, which is
released as an open-source platform\footnote{https://github.com/shebuxin/pfagent} for community-wide evaluation,
testing, and further development.

The rest of this paper is organized as follows.
Section~\ref{sec:agent} introduces the design philosophy of the power-system agent and details the technical framework.
Section~\ref{sec:pfagent} describes the self-evolution mechanism and the AI-assisted fixing loop.
Section~\ref{sec:eval} presents the evaluation method, including metrics and an ablation plan.
Section~\ref{sec:case} reports the case study on IEEE benchmark systems.
Section~\ref{sec:conclusion} concludes the paper and outlines future work.

% ========================================================================================
% Section II: Design Philosophy and Technical Framework
% ========================================================================================

\section{Power Flow Agents: Design Philosophy and Technical Framework}
\label{sec:agent}

\subsection{Design Philosophy}

A \emph{power-system agent} is defined here as a software system that accepts a natural-language study objective, converts it into a sequence of admissible simulator actions, executes those actions, and returns a
structured output.
Unlike prompt engineering-based question-answering systems, an agent must maintain execution state, enforce domain constraints, and produce outputs whose correctness can be checked independently of the agent itself.

This paper uses power flow tasks as the base implementation scope. 
Four properties guide the design of \emph{PFAgent}.
\begin{enumerate}
\item \textbf{Tractability.}
      Text-to-simulation is treated as a problem decomposed into named stages with structured outputs, including parsed intent, retrieved context, generated code, execution results, and repair traces.
      Each stage can be independently executed, inspected, edited, and re-run, so failures are localized to a specific stage rather than to the whole pipeline.
\item \textbf{Interactivity.}
      The agent supports multi-turn sessions.
      Earlier case modifications persist unless the user explicitly overrides them, and the user interface provides real-time access to code, execution logs, and error messages.
      Users can also step in at any point, including editing the generated code, re-running selected steps, or customizing plots.
\item \textbf{Reproducibility.}
      Every study output is backed by an executable Python script that runs against the same ANDES case in the same workspace. The result is a machine-parseable JSON structure that encodes all numerical claims.
\item \textbf{Self-evolution.}
      The agent is not static. Verified failures are attributed to specific root causes, and the resulting fixes are persisted as updates that take effect in subsequent sessions.
      Deployment feedback from users enters the same improvement path.
\end{enumerate}

\subsection{Technical Framework}
\label{subsec:framework}

\begin{figure*}[!t]
\centering
\includegraphics[width=0.85\linewidth]{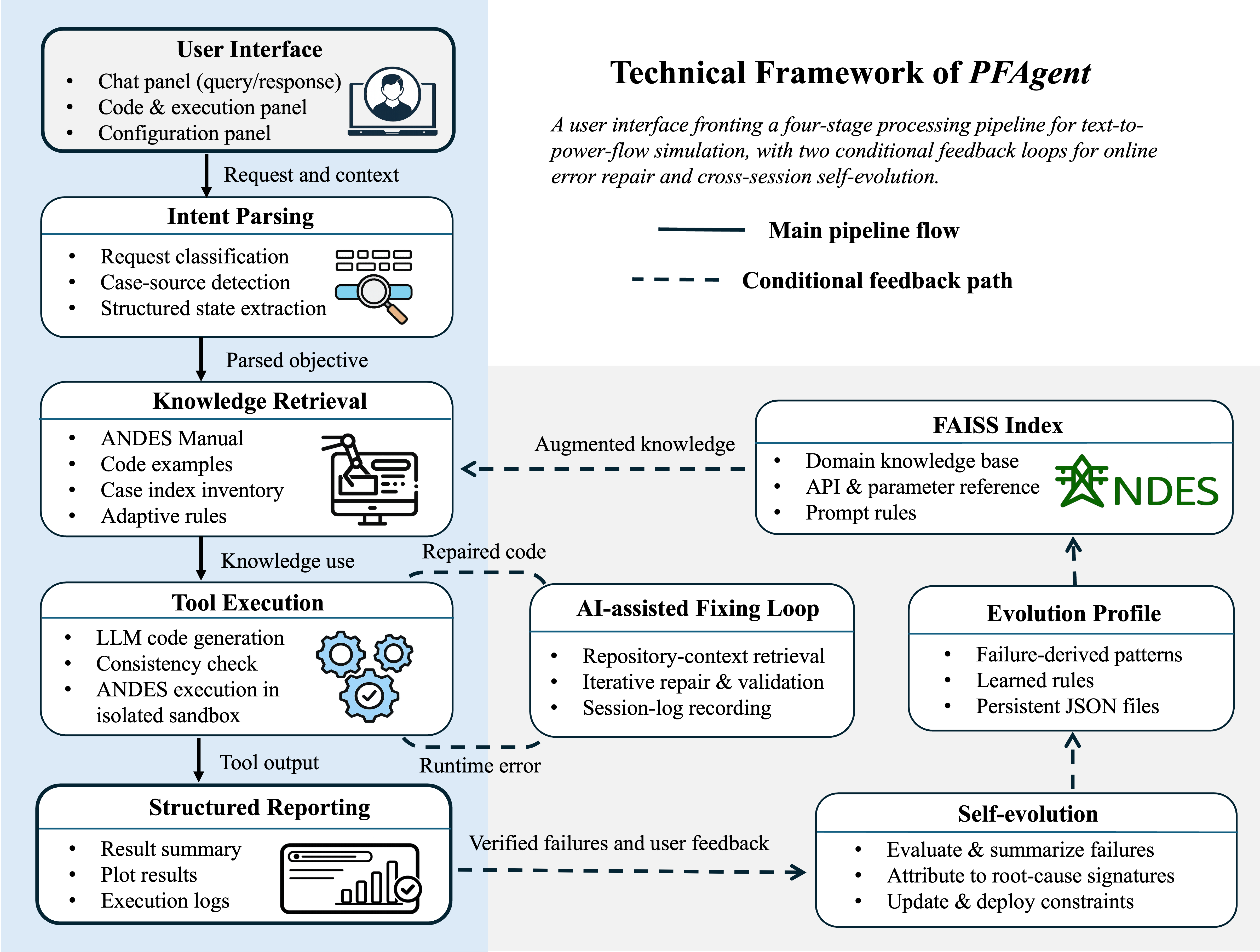}
\caption{Technical framework of \emph{PFAgent}:
the left column (blue) is the session query pipeline; the right column (gray) contains the feedback loops for error repair and self-evolution.}
\label{fig:architecture}
\end{figure*}

Fig.~\ref{fig:architecture} presents the overall architecture of \textit{PFAgent}, comprising five functional modules.
A user interface fronts a four-stage processing pipeline of intent parsing, knowledge retrieval, tool execution, and structured reporting.
The user interface includes a setting panel for configuration, a chat panel for natural-language queries and responses, a code-and-execution panel for online inspection, and a plot viewer for generated figures and logs. The delivered user interface is presented in Section~\ref{sec:case}.

Between modules, there are data flowing along the main pipeline through solid arrows.
The user interface passes the request and conversational context to the intent parser, which produces a parsed objective to the
knowledge-retrieval stage.
The retrieval stage augments the prompt with domain knowledge from a Facebook AI Similarity Search (FAISS)-indexed ANDES manual, curated code examples, a live case-index inventory, and adaptive rules.
Then, it forwards the prompt to the tool-execution stage.
The execution results are packaged by the structured-reporting stage into a result summary, plot modules, and full execution logs.

Two conditional feedback paths (dashed arrows) complement the main
pipeline.
First, when a runtime error occurs during tool execution,
the AI-assisted fixing loop assembles the failing code, the error
trace, and repository-level context to produce a repaired script that re-enters the execution stage.
Second, verified failures and user feedback collected after reporting are routed to the self-evolution mechanism, which attributes failures to root-cause signatures, updates constraints, and passes the results to the evolution profile.
The updated profile feeds back into the FAISS index as adaptive rules, closing a cross-session improvement loop that requires no model retraining.

\subsection{Design of Functional Modules}
% Four functional modules in the main pipeline are described below.

\subsubsection{Intent Parsing}

The first stage classifies the incoming user message and extracts the structured information needed by downstream steps.
\begin{itemize}
\item Identify the request type among runnable code, conceptual explanation, and debugging insight. This classification is performed by pattern-based detectors that inspect the effective user context for markers such as ``runnable Python code'' or ``explain'' or ``why''.
\item Detect the case source. \emph{PFAgent} distinguishes built-in ANDES benchmark cases from user-uploaded files on the exact uploaded filename. 
Confusing these two paths is one of the most common causes of grounding failure in LLM-generated power-system code.
\item Extract the intent markers that specify the study objectives, such as voltage check, load scaling, setpoint adjustment, or N--1 analysis.
\end{itemize}

The output of this stage is a structured parsed objective that records the case reference, the device parameters, and the accumulated modifications from prior turns. 
When the coding gate is triggered, the downstream tool execution module bypasses the LLM entirely and produces a template-based Python script that is guaranteed to follow the correct ANDES workflow order.

\subsubsection{Knowledge Retrieval}

This stage assembles the context that will form the system
prompt for the LLM. Four sources of knowledge are combined.
\begin{itemize}
\item The ANDES manual is pre-indexed into a FAISS vector store using sentence-level embeddings.
At query time, the retriever selects the contiguous page windows of the manual that are most relevant to the current prompt.
This window-based strategy avoids the fragmentation that arises from small chunk retrieval, which is a common source of errors in code-generation tasks.
\item Code examples are loaded from a curated JSON file.
These examples demonstrate the expected code style, import order, and output format.
\item The case index inventory is built by loading the
active ANDES case and rendering the real index strings together with their bus connectivity.
This eliminates the class of errors in which the LLM guesses a device index instead of using the actual string identifier.
\item Adaptive rules are loaded from the evolution profile.
These are prompt-level rules that encode lessons learned from prior verified failures.
The set is updated by the self-evolution mechanism described in Section~\ref{sec:pfagent}.
\end{itemize}

\subsubsection{Tool Execution}

This stage converts the prompt into executable Python code
and runs it against the ANDES simulator. The process has three phases.
\begin{itemize}
\item Code Generation: The system prompt and the conversation history are sent to the LLM.
The response is normalized to ensure it contains Python code block, includes necessary imports, and ends with a output statement.
\item Code Inspection: Before execution, the generated code undergoes two static checks.
An Abstract Syntax Tree (AST)-level syntax check catches malformed Python code.
A pattern-based case validator ensures that built-in and uploaded cases are loaded correctly and that device
indices are resolved correctly.
If either check fails, the error message is fed back to the LLM as a compilation-error signal, and the LLM is asked to regenerate.
This retry loop runs up to a configurable number of attempts.
\item Code Execution: Valid code is executed in an isolated environment.
The executor captures simulation results and generated plot files.
If the code raises an exception during ANDES execution, the error is surfaced to the user together with a ``Fix Error with AI'' action, which triggers the AI-assisted fixing
loop described in Section~\ref{sec:pfagent}.
\end{itemize}

\subsubsection{Structured Reporting}

The final stage packages the execution output into a form that is both user-readable and machine-checkable.
The structured output is a JSON object printed by the generated script at the end of execution.
Its keys are prompt-specific. For example, a voltage-ranking task includes bus index, voltage magnitude, and rank.
Its values are the numbers computed by ANDES.
This design serves two purposes: it gives the user a concise summary of the study result, and it gives the evaluator a structured output to score.
Plot files are captured from the session workspace and rendered in the user interface alongside the conversational response.
The full execution log, including the generated code, runtime output, and any error-fix history, is recorded to a session log file.
This log ensures that every claim made by the agent can be traced back to a specific script and a specific ANDES call.

% ========================================================================================
% Section III: Self-evolution and AI-assisted Fixing
% ========================================================================================

\section{Self-evolution and AI-assisted Fixing Mechanism}
\label{sec:pfagent}

This section introduces two mechanisms that close the loop between agent execution and agent improvement.
The self-evolution mechanism operates across sessions. It uses verified failures from benchmark runs and user feedback to update the agent's prompt rules for all future sessions.
The AI-assisted fixing loop operates within a single session.
When generated code fails at runtime, it invokes a separate repair model with the conversational context, the failing code, the execution error, and repository-level knowledge to produce a fix online.

\subsection{Self-evolution Mechanism}
\label{subsec:selfevo}

A power-system agent inevitably encounters failures that could not have been anticipated at initial design time. Unseen natural-language phrasings, undocumented edge cases, or format inconsistencies can result in unexpected errors.
If these failures are only addressed by re-prompting or manual code patches, the agent remains static and the same class of error recurs in every new session.

\begin{figure}[!t]
\centering
\includegraphics[width=\linewidth]{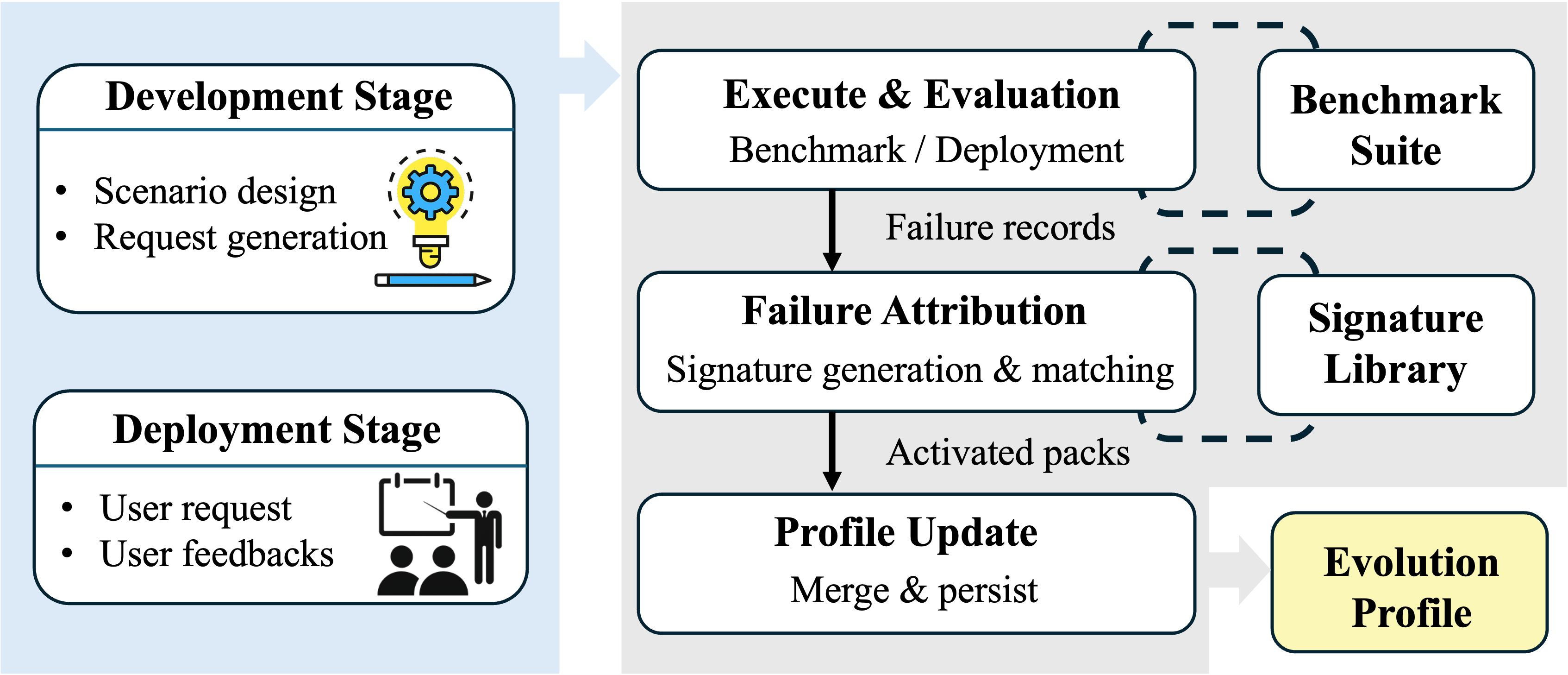}
\caption{Self-evolution mechanism.}
\label{fig:selfevo}
\end{figure}

\emph{PFAgent} addresses this with a self-evolution mechanism that operates in two complementary stages, as shown in
Fig.~\ref{fig:selfevo}.
In the development stage, the agent is exercised against a generated benchmark suite before release; in the deployment stage, the same improvement loop continues with real user requests and feedback collected through the user interface.
Both stages share a common three-phase processing pipeline, including execute and evaluate, failure attribution, and profile update. In this way, lessons learned in either stage accumulate in the same evolution profile.

\subsubsection{Development Stage}

Before the agent is released, a deterministic benchmark suite is constructed by emulating realistic user behavior.
Each scenario in the suite encodes a study session.
The scenario design step selects case families, source configurations, and follow-up operation types.
The request generation step converts each scenario specification into natural-language prompts that mimic the phrasing diversity of real users.

The agent is then run on the full suite.
Every turn produces generated code, execution outputs, and a structured output in JSON format.
The evaluation module scores each turn on six dimensions introduced in Section~\ref{sec:eval}, including format, grounding, continuity, execution, semantic correctness, and output quality.
This yields a per-turn pass/fail label together with a list of specific failure categories.
The resulting failure records enter the shared processing pipeline described below.

\subsubsection{Deployment Stage}

After release, the same structured logging pipeline records every
production session.
Each chat turn, code execution event, and user-initiated feedback action is written to a per-session JSON log and appended to a global event stream.
These records supplement the development-stage benchmark failures with real-world usage patterns that the benchmark may not cover, such as novel phrasing, unexpected case files, or domain-specific conventions from individual users.
A feedback panel integrated into the user interface allows users to flag errors and annotate root causes directly.

\subsubsection{Failure Attribution}

Both stages feed failure records into a common attribution step.
Each record is matched against a library of failure signatures.
A signature is a triple $(\mathcal{P}, \mathcal{E}, \mathcal{I})$ where $\mathcal{P}$ is a set of prompt patterns, $\mathcal{E}$ is a set of execution-error patterns, and $\mathcal{I}$ is a set of human issue patterns.
A failure record activates a signature if it matches any pattern in at least one of the three sets.

Each signature is linked to one or more constraint packs.
A constraint pack contains three components:
(i)~prompt guidance sentences that are injected into the system prompt at query time,
(ii)~pattern overrides that extend the intent parser's expression vocabulary,
and
(iii)~marker overrides that extend the set of natural-language phrases recognized by the structured code-generation gate.

% TBD: Converter to an algorithm table?

% For example, the signature \texttt{device\_idx\_cast\_to\_int} fires
% when an execution error matches
% \texttt{invalid literal for int() with base 10: 'PQ\_*'}. It
% activates the constraint pack
% \texttt{string\_device\_idx\_constraint}, whose prompt guidance
% instructs the LLM to preserve string-valued ANDES device identifiers
% instead of forcing integer conversion. Similarly, the signature
% \texttt{corridor\_outage\_language} fires when the prompt mentions
% ``corridor'' or ``out of service'' and the execution output contains
% \texttt{KeyError: 'in\_service'}. It activates both
% \texttt{corridor\_outage\_aliases} (which adds line-outage phrasings
% to the intent parser) and \texttt{line\_outage\_api\_constraint}
% (which instructs the LLM to use the correct ANDES outage API). The
% current library contains eight failure signatures covering device
% index resolution, percentage-based demand scaling, regulator-target
% phrasing, corridor-outage language, N--1 screening-set language,
% non-runnable response contracts, and follow-up continuity loss.

\subsubsection{Profile Update}

The attributed failures are aggregated into an evolution profile. 
It is a JSON file that records the active constraint packs, the accumulated prompt guidance, the extended pattern and marker overrides, and a root-cause summary with example turns.
Profiles from different sources can be merged.
The merge operation takes the union of all active packs, deduplicates guidance strings, and accumulates root-cause counts.
The merged profile is saved and loaded by the knowledge-retrieval stage at the start of every new session.
This ensures that lessons learned from either the development or the deployment stage take effect immediately in all subsequent sessions.

\subsection{AI-Assisted Fixing Loop}
\label{subsec:aifixer}

\begin{figure}[!t]
\centering
\includegraphics[width=0.8\linewidth]{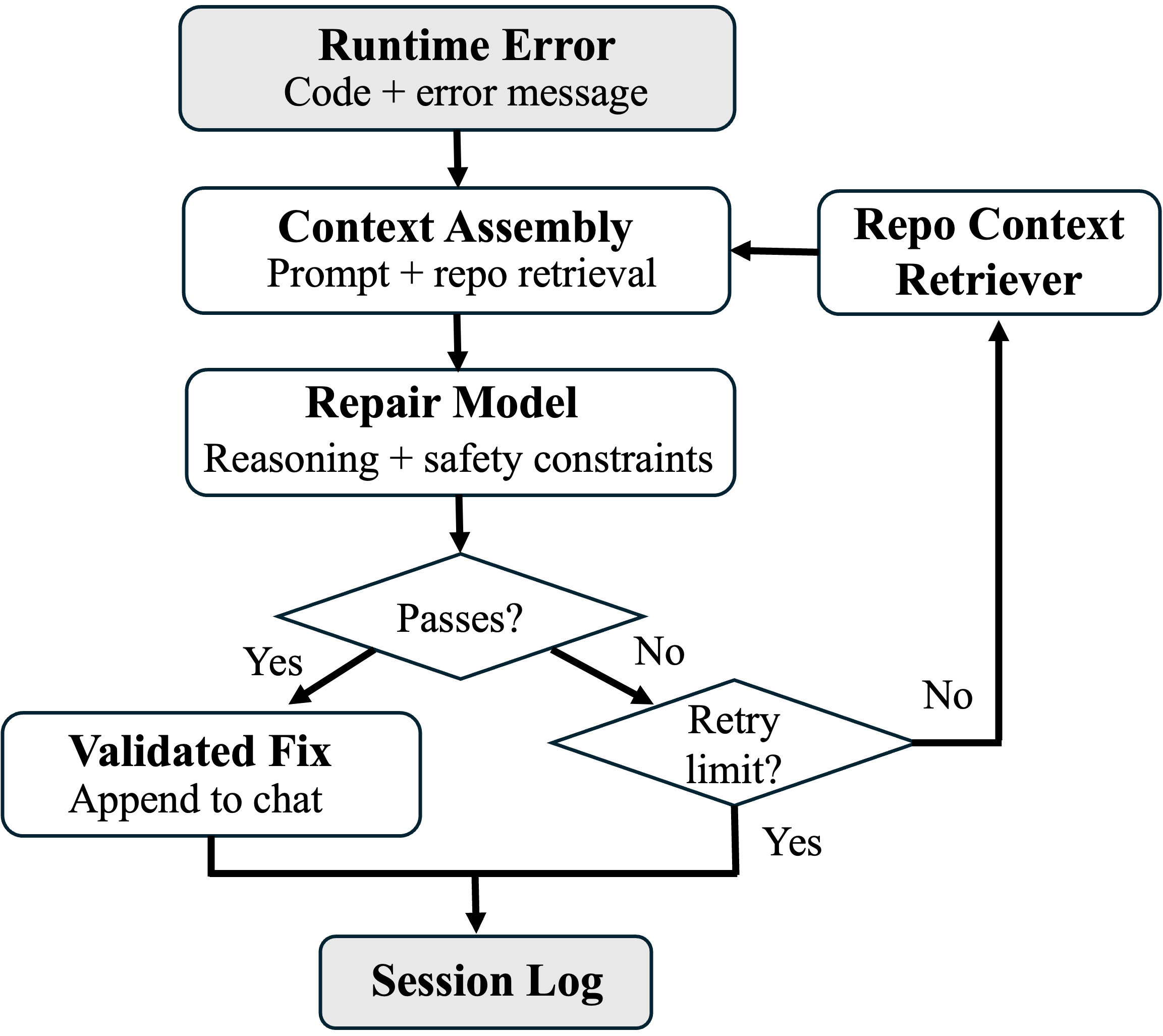}
\caption{AI-assisted fixing loop.}
\label{fig:aifixer}
\end{figure}

The self-evolution mechanism improves the agent over time, but it does not help a user whose code has already failed in the current session.
For this case, \emph{PFAgent} provides a built-in AI-assisted fixing loop that operates online within the same conversation.
As shown in Fig.~\ref{fig:aifixer}, when a generated Python script raises an exception during ANDES runtime execution, the user interface surfaces the error output together with a ``Fix Error with AI'' action. Clicking this action triggers a repair request, which is processed as follows.

\subsubsection{Context Assembly}

The repair pipeline first assembles a comprehensive fix prompt that combines six sources of context:
(i)~the original user message that triggered the failing code,
(ii)~the agent's response containing the failing code,
(iii)~the exact Python code block that was executed,
(iv)~the full output and error messages,
(v)~the list of files currently available in the session workspace, and
(vi)~the active ANDES case identifier and configurations.

In addition, a repository context retriever indexes the \emph{PFAgent} source tree, including application code, tests, examples, documentation, and verification scripts. This is done by splitting the content into overlapping text chunks.
At repair time, it extracts signal terms from the error message and the failing code, and retrieves the top-$k$ repository items whose paths or contents best match those terms.
This allows the fixer to reference actual ANDES API usage patterns, constraint implementations, and test
fixtures, rather than relying on the repair model's general knowledge.

\subsubsection{Repair and Validation}

The assembled prompt is sent to a dedicated repair model, which is a reasoning model separate from the main chat model.
If local validation is enabled, the repaired code is immediately executed in the same session workspace.
If execution still produces an error, the new error output replaces the old one in the fix prompt, and the repair model is called again.
This iterative cycle continues until the code executes without error or a configurable retry limit is reached.

\subsubsection{Termination and Recording}

On success, the validated fix is appended to the conversation history with a note indicating how many repair iterations were needed and whether local validation passed.
On failure, the best available fix is presented together with the remaining error output, so the user can diagnose the issue manually.
Regardless of outcome, the full repair event is recorded in the session log through the same feedback pipeline used by
the self-evolution mechanism.
If the repair failure matches one of the known failure signatures, the corresponding constraint pack is queued for activation in the next profile update cycle.
This design ensures that the fixing loop and the self-evolution mechanism reinforce each other.

% ========================================================================================
% Section IV: Evaluation Metrics and Framework
% ========================================================================================

\section{Evaluation Metrics and Framework}
\label{sec:eval}

This section defines the metrics used to evaluate \emph{PFAgent} and describes the ablation plan for isolating the contribution of fine-tuning and RAG techniques.

\subsection{Metrics}
\label{subsec:metrics}

A single task in the \emph{PFAgent} benchmark is a multi-turn conversation scenario, such as a base-case power flow study and cumulative case modifications.
Each turn is scored individually on six dimensions, and the conversation score is the mean of its turn scores.
The six dimensions and their point allocations are as follows.

\subsubsection{Format (10 points)}
As shown in \eqref{eq:format}, this dimension checks whether the response contains exactly one fenced Python code block.
If the response omits code, includes conflicting scripts, or returns results without executable content when code was requested, the format score is zero.
\begin{equation}
S_{\mathrm{fmt}} =
  \begin{cases}
    10, & \text{if exactly one fenced Python block},\\
    0,  & \text{otherwise}.
  \end{cases}
\label{eq:format}
\end{equation}

\subsubsection{Grounding (25 points)}

This dimension checks whether the generated code uses the correct ANDES API calls.
Each scenario turn specifies a set of code checks, including weighted expression patterns that must appear in the generated script.
Examples include the correct \texttt{andes.get\_case} call for built-in cases, the correct \texttt{andes.load} call for uploaded cases, the correct index set for line outages, and the expected output format.
A set of forbidden patterns is also checked, and any match results in a penalty.

Let $\{(w_j, p_j)\}_{j=1}^{J}$ be the weighted code checks for a given turn, where $w_j$ is the weight and $p_j$ is a binary indicator for whether pattern $j$ appears in the code. The grounding score is:
\begin{equation}
S_{\mathrm{gnd}} = 25 \times
  \frac{\sum_{j=1}^{J} w_j \cdot p_j}
       {\sum_{j=1}^{J} w_j}.
\label{eq:grounding}
\end{equation}

\subsubsection{Continuity (15 points)}

Multi-turn studies require that earlier modifications persist in later turns unless the user explicitly overrides them.
Each follow-up turn specifies carry-forward checks.
The weighted patterns must appear in the generated code to confirm that prior edits such as a load scaling or a PV setpoint change from a previous turn are still present.
The scoring formula is identical to~\eqref{eq:grounding}, scaled to 15 points:
\begin{equation}
S_{\mathrm{cont}} = 15 \times
  \frac{\sum_{j=1}^{J'} w_j \cdot p_j}
       {\sum_{j=1}^{J'} w_j}.
\label{eq:continuity}
\end{equation}

\subsubsection{Execution (20 points)}

This dimension checks whether the generated code runs to completion in the ANDES runtime without raising a Python exception.
The code is executed in an isolated sandbox with a code prefix that redirects \texttt{matplotlib} output to files.
The criterion is binary:
\begin{equation}
S_{\mathrm{exec}} =
  \begin{cases}
    20, & \text{if exit code } = 0,\\
    0,  & \text{otherwise}.
  \end{cases}
\label{eq:execution}
\end{equation}

Note that a zero execution score does not necessarily imply wrong reasoning.
It may result from a missing import, a type mismatch in a device index, or an ANDES API call that is valid but incompatible with the current version.

\subsubsection{Semantic Consistency (25 points)}

This dimension compares the output against an expected result computed independently by the verification module.
The verification module loads the same ANDES case, applies the same cumulative operations using ANDES API calls, runs power flow, and extracts the expected key--value pairs.

Let $\{k_1, \ldots, k_K\}$ be the keys specified by the verification module for a
given turn. For each key $k_i$, the expected value $v_i^{*}$ and the agent-produced value $v_i$ are compared.
Numeric values are compared with an absolute tolerance of $10^{-4}$; string and integer values are compared exactly. The semantic score is:
\begin{equation}
S_{\mathrm{sem}} = 25 \times
  \frac{\bigl|\{i : |v_i - v_i^{*}| \le 10^{-4}\}\bigr|}{K}.
\label{eq:semantic}
\end{equation}

\subsubsection{Output Quality (5 points)}

For turns that request a plot, the evaluation checks whether (a)~the expected plot filename appears in the \texttt{RESULT\_JSON}, and (b)~the corresponding image file was
created in the output directory:
\begin{equation}
S_{\mathrm{art}} =
  \begin{cases}
    5, & \text{if plot file is produced and correctly named},\\
    0, & \text{otherwise (or if no plot was requested)}.
  \end{cases}
\label{eq:artifact}
\end{equation}

\subsubsection{Aggregation}

The total turn score is:
\begin{equation}
S_{\mathrm{turn}} = S_{\mathrm{fmt}} + S_{\mathrm{gnd}}
  + S_{\mathrm{cont}} + S_{\mathrm{exec}} + S_{\mathrm{sem}}
  + S_{\mathrm{art}},
\label{eq:total}
\end{equation}
with a maximum of 100. A turn passes if and only if it receives full marks in all six dimensions.
The conversation score is the mean of the three turn scores, and a scenario passes only if all three turns pass.

\subsection{Ablation Plan}
\label{subsec:ablation}

The evaluation framework supports four agent modes that form a natural ablation sequence.
\begin{enumerate}
\item \textbf{Base OpenAI.}
      The base LLM receives the user prompt without retrieval, without fine-tuning, and without ANDES-specific constraints.
      This mode serves as the lower-bound baseline, measuring how far a general-purpose LLM can go when given only the task description.
\item \textbf{Fine-tuned.}
      A fine-tuned variant of the same base model, trained on ANDES code examples, LLM-generated prompt/code pairs, hand-curated hard cases, and verified single- and multi-turn conversations.
      The pooled samples are split 90/10 into train/validation JSONL files.
      This mode isolates the effect of domain-specific training data without retrieval.
\item \textbf{RAG.}
      The base LLM augmented with the full knowledge-retrieval pipeline and the structured codegen gate.
      The retrieval layer indexes the ANDES manual into a FAISS vector store with window-level passages, and adds a library of example codes as few-shot context.
      This mode isolates the effect of retrieval and prompt engineering without fine-tuning.
\item \textbf{Fine-tuned + RAG.}
      The fine-tuned model combined with the full retrieval pipeline.
      This is the recommended production mode and the upper bound of the ablation.
\end{enumerate}

% Comparing \emph{Base} vs.\ \emph{Fine-tuned} isolates the effect of
% training data. Comparing \emph{Base} vs.\ \emph{RAG} isolates the
% effect of the retrieval and prompt-engineering pipeline. Comparing
% \emph{RAG} vs.\ \emph{Fine-tuned + RAG} reveals whether fine-tuning
% provides incremental benefit on top of retrieval.

% Each mode is evaluated on the same deterministic scenario suite, using
% the same oracle and the same scoring rubric. The scenarios cover four
% IEEE benchmark case families (IEEE~14, IEEE~39, Kundur, PJM~5) in
% both built-in and uploaded configurations, with follow-up operations
% including load addition, load scaling, PV and slack setpoint
% adjustment, line outage, N--1 contingency screening, and
% voltage/angle ranking with threshold checks.

% To further isolate the contribution of the self-evolution mechanism, the evaluation includes a before/after comparison.
% Specifically, the same scenario suite is run before and after the evolution profile is updated from a prior round of failure attribution. 

% The 100-scenario suite serves as the early-stage baseline, while the expanded 164-scenario suite with additional open-ended tasks serves as the post-evolution evaluation.
% This comparison directly measures the improvement attributable to guardrail updates derived from verified failures, independent of any LLM retraining.

% ========================================================================================
% Section V: Case Study
% ========================================================================================

\section{Case Study}
\label{sec:case}

This section introduces the delivered \emph{PFAgent} and evaluates its performance on representative power-flow study tasks.

\subsection{Delivered Platform}

The four-stage pipeline described above is delivered through a Streamlit-based web interface.
Additional technical details, fine-tuning and RAG data, and training and verification logs can be found in the GitHub repository~\cite{she2026pfagent}.

\begin{figure*}[!t]
\centering
\includegraphics[width=1\linewidth]{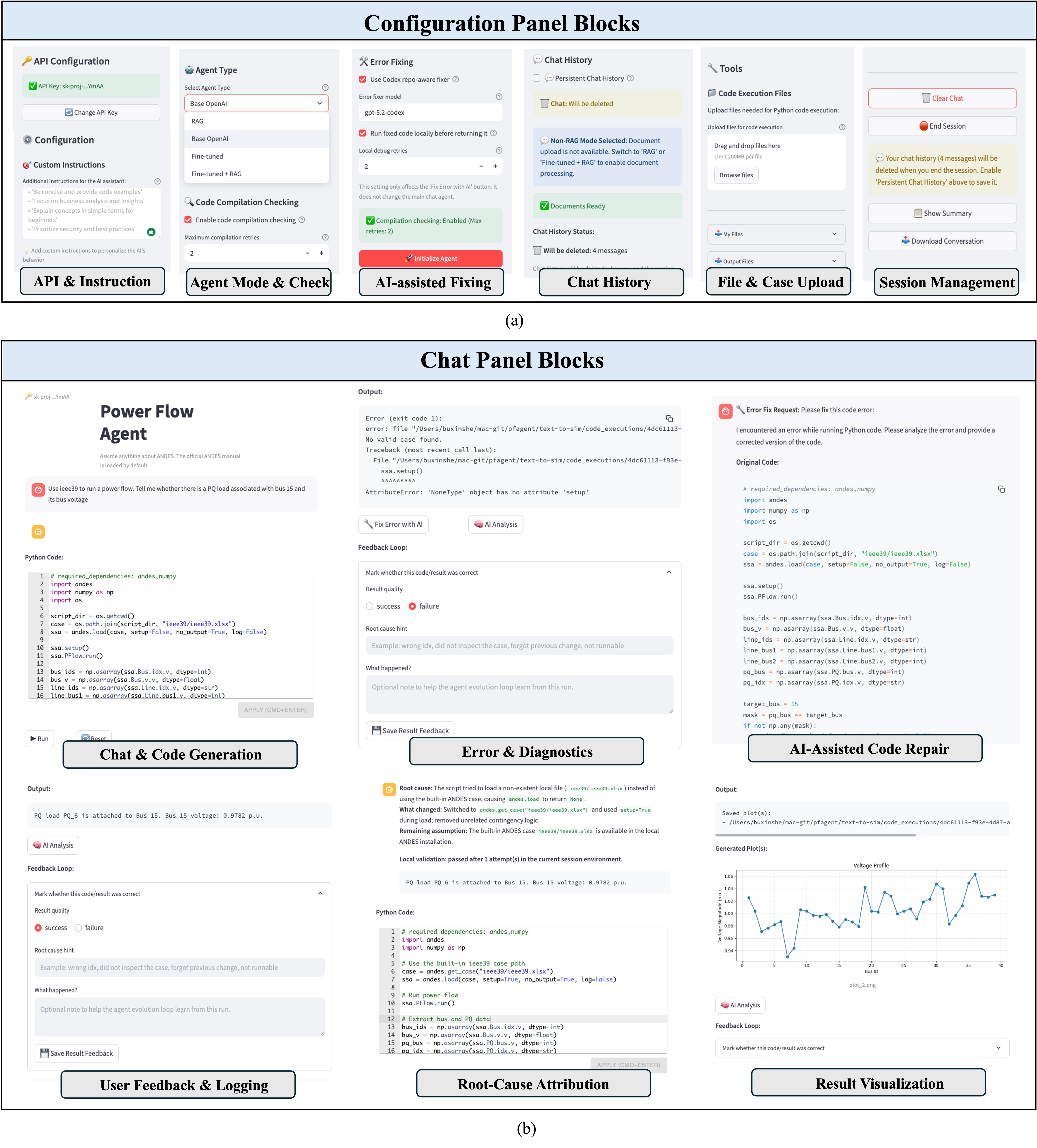}
\caption{User interface of \emph{PFAgent}:
(a)~Configuration panels;
(b)~Chat panels.}
\label{fig:ui}
\end{figure*}

Fig.~\ref{fig:ui} shows the two main panels of the interface. 
The configuration panel provides sidebar controls for API key entry and prompt customization, agent-mode selection, with validation toggles, AI-assisted fixing options, conversation history management, case-file upload, and session persistence.
The chat panel hosts the interactive study session, comprising a conversational code-generation area, an error diagnostics view, an AI-assisted code-repair interface, a user-feedback and logging module that feeds the self-evolution mechanism, a root-cause attribution display, and an inline result-visualization panel for plots and output.

\subsection{Study Setup}
\label{subsec:setup}

\subsubsection{Benchmark Suite}

The evaluation uses a deterministic scenario suite generated for demonstration.
Two suite sizes are reported:
an initial 100-scenario suite used during early development, and an expanded 164-scenario suite that adds more challenging tasks introduced after the first round of self-evolution.
Each scenario consists of three turns, including a baseline request (Turn~1), a first follow-up with a case modification (Turn~2), and a second follow-up with an additional modification (Turn~3).

\subsubsection{Case Families and Sources}

Four IEEE benchmark case families are used: IEEE~14, IEEE~39, Kundur, and PJM~5.
Each family appears in two source configurations, including built-in cases loaded from the ANDES package via \texttt{andes.get\_case} and user-defined cases loaded via \texttt{andes.load}.

\subsubsection{Task Types}

The follow-up modifications cover a cross-section of power-flow analysis tasks, including load addition before setup, uniform scaling after setup, slack-bus and PV-bus voltage setpoint adjustment, targeted load modification at a specific bus, targeted generation change at a specific bus, line outage by bus pair, N--1 contingency analysis over a candidate line list, voltage and line-angle ranking with threshold filtering, and voltage-profile plot generation.

\subsubsection{Agent Modes}

Four modes introduced in Section~\ref{subsec:ablation} are evaluated and compared on the benchmark suite under the same settings.

\subsubsection{Self-Evolution Comparison}

To measure the effect of the self-evolution mechanism, the Fine-tuned~+~RAG mode is evaluated on the 164-scenario suite both before and after the evolution profile is updated from the failure-attribution pass.
The pre-evolution run uses the initial prompt template, while the post-evolution run uses
the updated constraint packs derived from the first round of verified failures.

\subsection{Results}
\label{subsec:results}

\subsubsection{100-Scenario Benchmark}

\begin{figure*}[!t]
\centering
\includegraphics[width=0.32\textwidth]{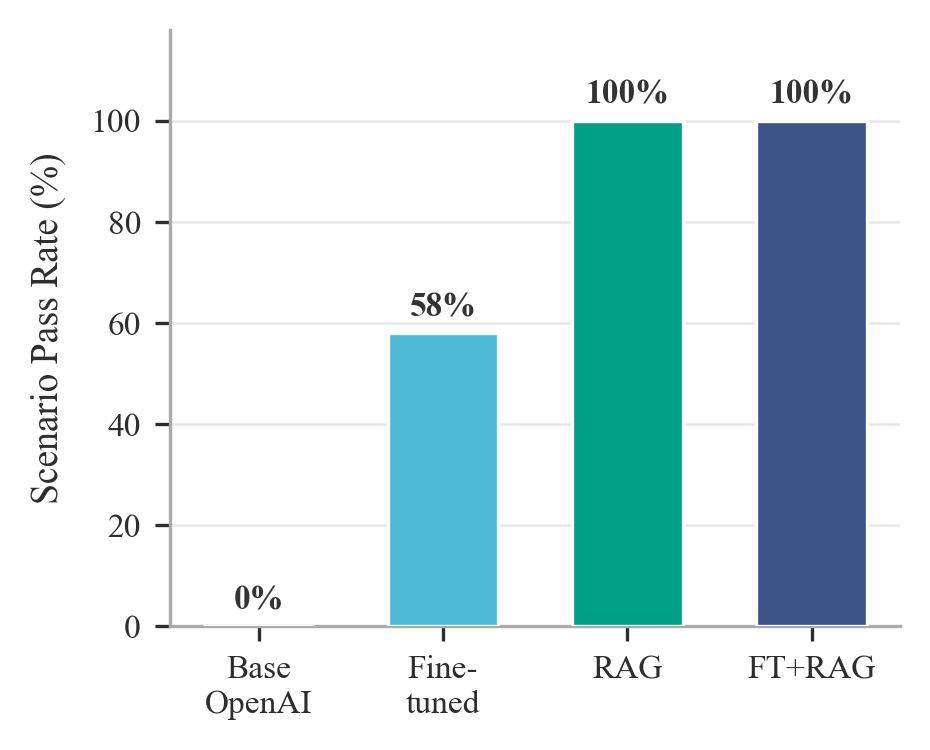}%
\hfill
\includegraphics[width=0.34\textwidth]{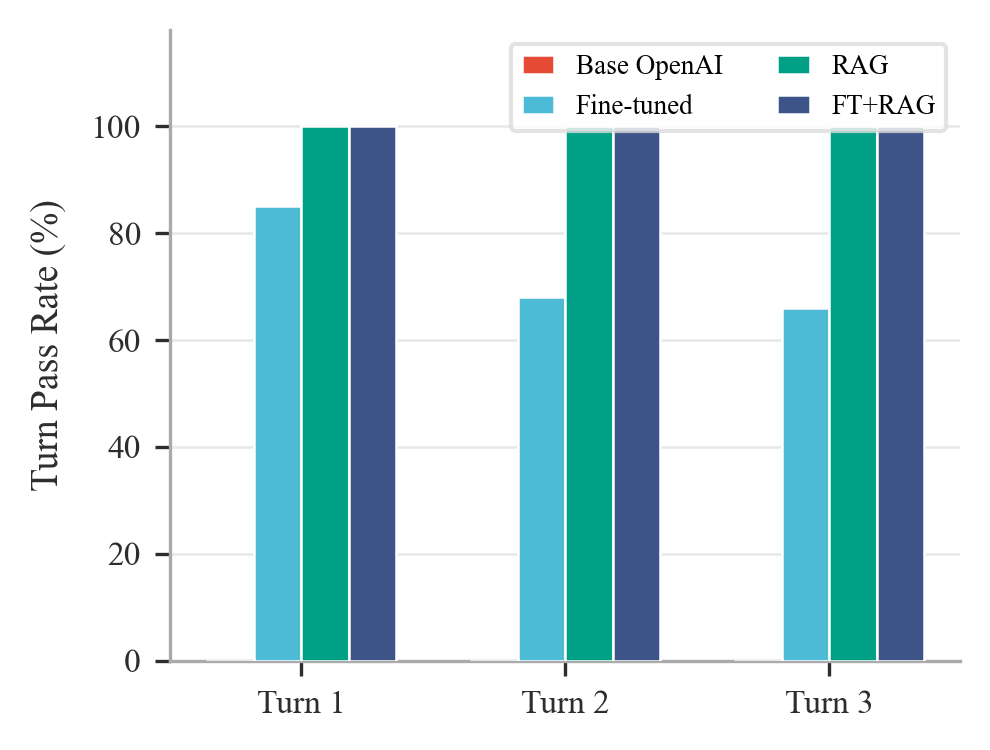}%
\hfill
\includegraphics[width=0.34\textwidth]{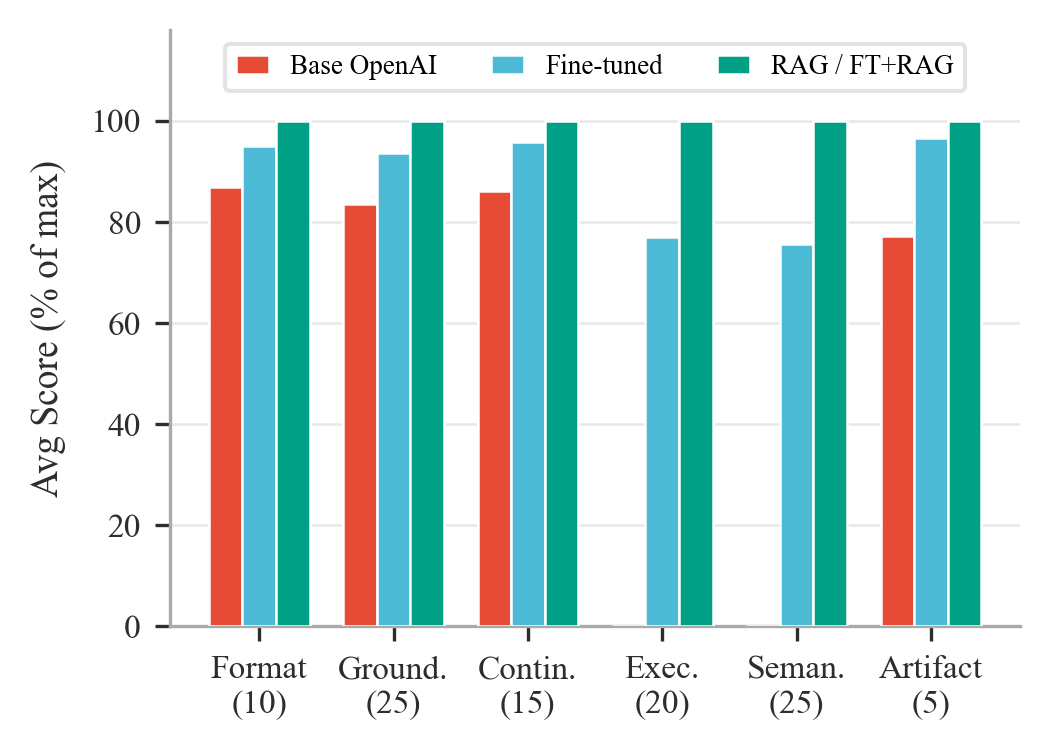}
\caption{100-scenario benchmark results:
(a)~Scenario pass rate by mode.
(b)~Turn-level pass rate;
(c)~Dimension-level performance comparison.}
\label{fig:results_100}
\end{figure*}

Fig.~\ref{fig:results_100}(a) visualizes the metrics result for the 100-scenario benchmark.
The retrieval-based modes both achieve a 100\% scenario pass rate and a perfect average conversation score of 100.0, meaning every turn in every scenario received full marks on all six evaluation dimensions.
The Fine-tuned mode without retrieval reaches 58\% scenario pass rate with an average score of 86.44.
The Base~OpenAI mode, which lacks both retrieval and domain tuning, does not pass any scenario and averages 46.35.

\subsubsection{Turn-Level Analysis}

Fig.~\ref{fig:results_100}(b) shows the turn-level breakdown.
Two patterns are visible.
First, the Fine-tuned mode degrades across turns: Turn~1 pass rate is 85\%, Turn~2 drops to 68\%, and Turn~3 drops further to 66\%.
This indicates that the fine-tuned model struggles to preserve cumulative modifications in later turns without the explicit continuity support provided by the retrieval pipeline.
Second, the retrieval-based modes show no degradation at all.
Their per-turn pass rate is 100\% for all three turns.

\subsubsection{Dimension-Level Analysis}

Fig.~\ref{fig:results_100}(c) shows the average score as a percentage of the maximum for each evaluation dimension.
The Base~OpenAI mode scores well on format (87\%) and reasonably on grounding (84\%) and continuity (86\%), meaning the LLM usually produces a single code block with roughly correct API structure.
However, it fails completely on execution (0\%) and semantics (0\%). The generated code does not run, and consequently no JSON output is produced.
The Fine-tuned mode improves execution to 77\% and semantics to 76\%, but these remain the weakest dimensions.
The retrieval-backed modes achieve 100\% on all six dimensions.

\subsubsection{Failure Analysis}

\begin{figure}[!t]
\centering
\includegraphics[width=0.48\textwidth]{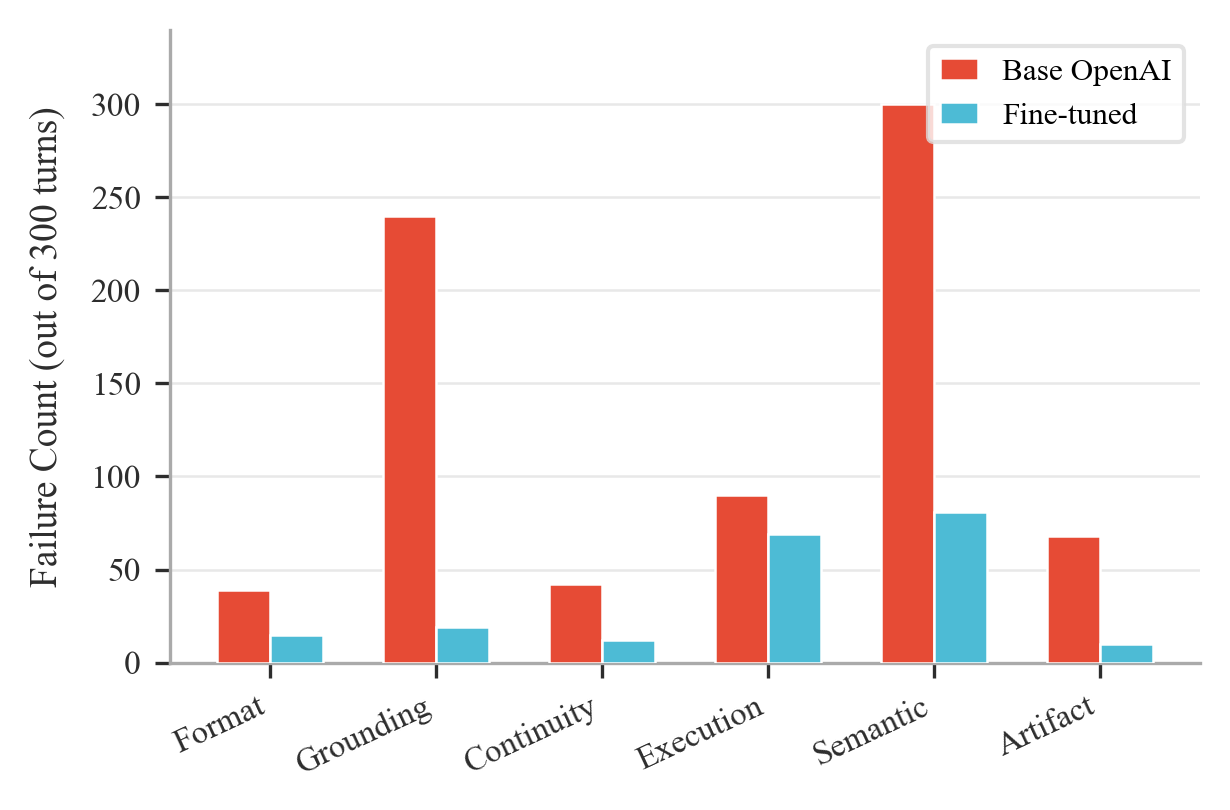}
\caption{Failure-category distribution for Base~OpenAI and
Fine-tuned modes on the 100-scenario benchmark.}
\label{fig:failure_family}
\end{figure}

Fig.~\ref{fig:failure_family} shows the failure-category distribution for the Base~OpenAI and Fine-tuned modes. RAG and Fine-tuned~+~RAG have zero failures, so they are omitted.
For Base~OpenAI, all 300 turns fail on execution and semantics. The grounding failures account for 240 out of 300 turns, indicating widespread API misuse.
For Fine-tuned, the dominant failure categories shift to semantics (81) and execution (69). It means the model produces code that is closer to correct but still contains errors that prevent convergence or produce wrong numerical results.

\subsubsection{Family-Level Analysis}

\begin{figure}[!t]
\centering
\includegraphics[width=0.48\textwidth]{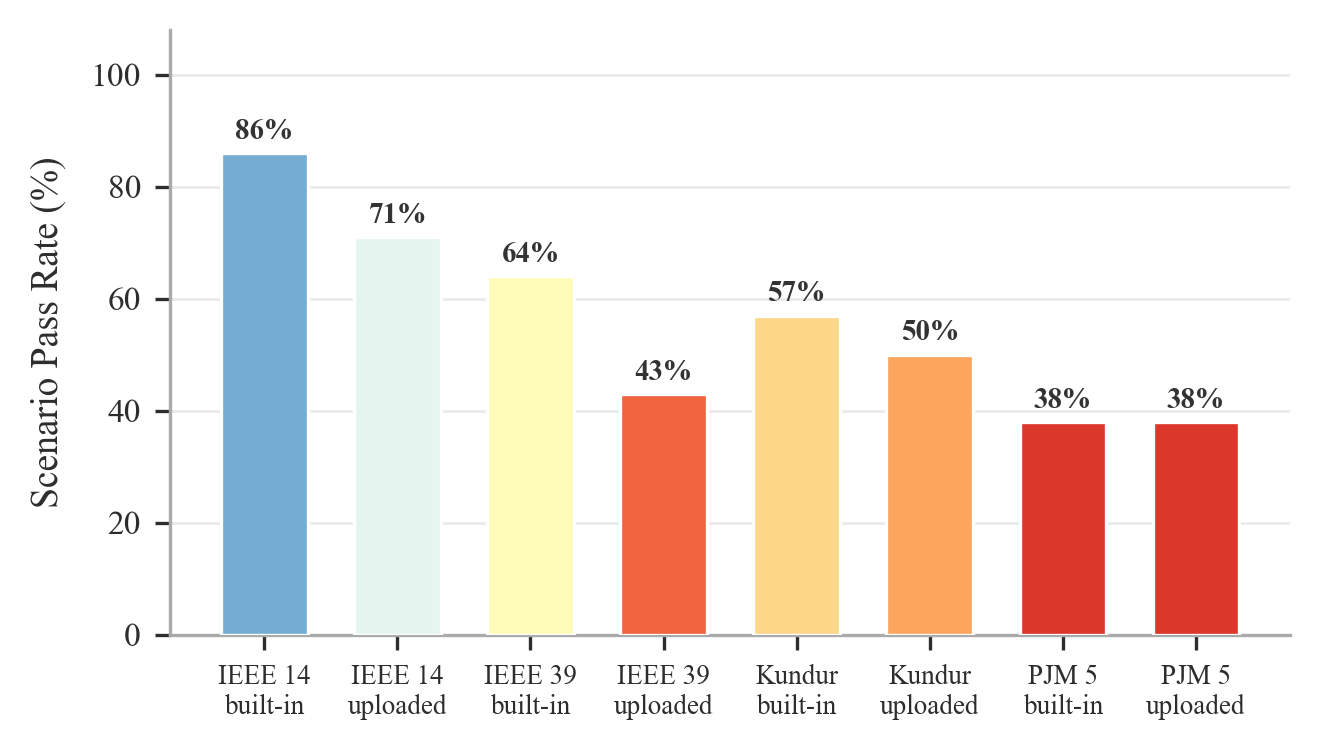}
\caption{Family-level scenario pass rate for the Fine-tuned mode.}
\label{fig:family_ft}
\end{figure}

For the retrieval-based modes, all eight cases achieve 100\% scenario pass rate.
For the Fine-tuned mode, performance varies by family and source.
Fig.~\ref{fig:family_ft} shows that built-in IEEE~14 reaches 85.7\% while PJM~5 (both sources) reaches only 37.5\%.
The uploaded variants consistently score lower than their built-in counterparts, confirming that case-source fidelity is a general challenge when retrieval and explicit loading rules are absent.

\subsubsection{Self-Evolution Effect}

\begin{figure}[!t]
\centering
\includegraphics[width=0.48\textwidth]{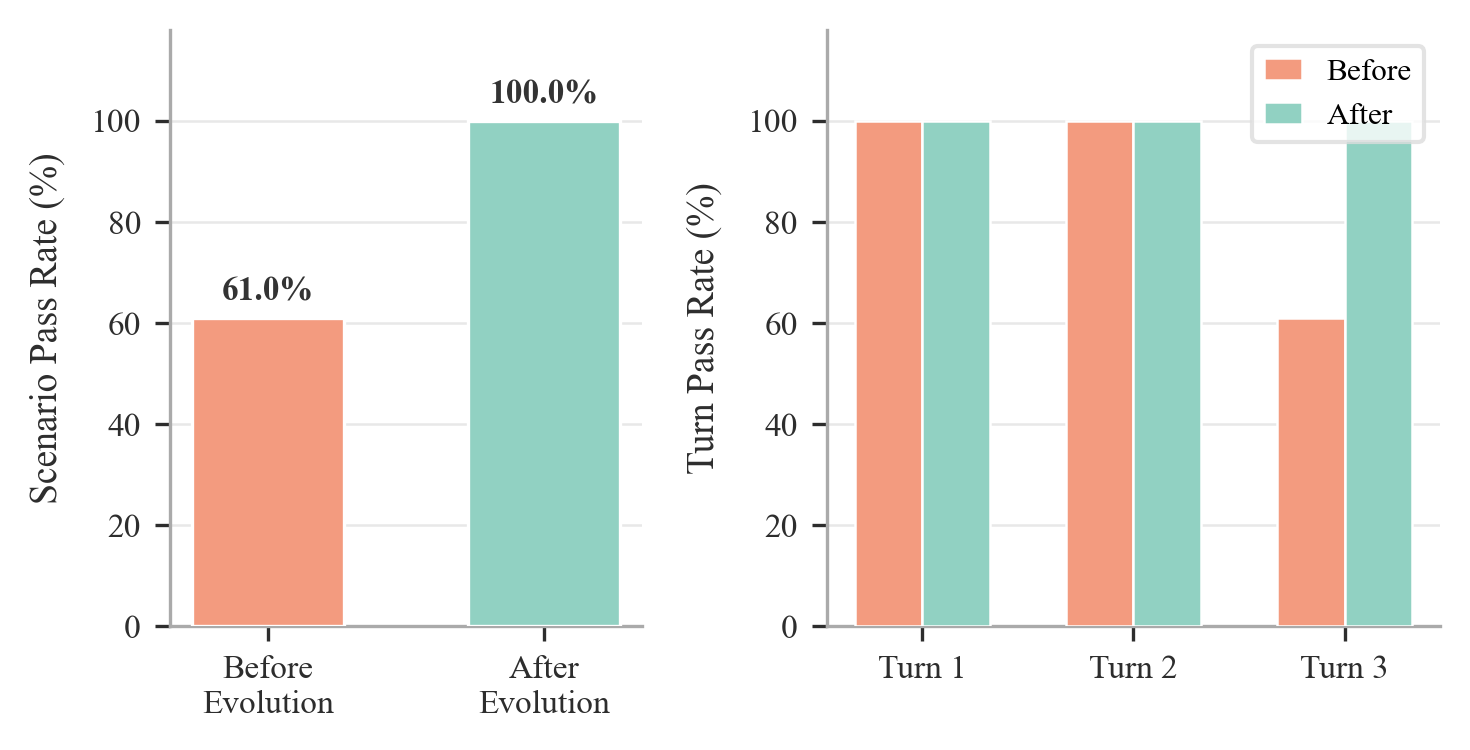}
\caption{Self-evolution before/after comparison on the 164-scenario
benchmark (Fine-tuned~+~RAG):
(a)~Scenario pass rate;
(b)~Turn-level analysis.}
\label{fig:evolution}
\end{figure}

Fig.~\ref{fig:evolution} reports the before/after comparison on the expanded 164-scenario suite.
Before the evolution update, the Fine-tuned~+~RAG mode achieves a 60.98\% scenario pass rate with an average score of 99.14.
Turns~1 and~2 pass at 100\%, but Turn~3 drops to 60.98\%.
Failure analysis reveals that nearly all Turn~3 failures are grounding failures concentrated in contingency tasks, where the agent uses an API form that is semantically correct but does not match the grounding rules.

After updating the evolution profile with new adaptive constraints targeting line-outage API usage and islanding detection, the same mode is rerun on the same 164-scenario suite.
The scenario pass rate recovers to 100.0\%, and all 492 turns pass with a perfect score.
No model retraining is involved.
The improvement comes entirely from the updated constraint packs injected into the system prompt.

\subsection{Discussion}
\label{subsec:discussion}

\subsubsection{Retrieval as the Dominant Factor}

Retrieval-based modes achieve 100\% pass rate on both the 100-scenario and the post-evolution 164-scenario suites, while fine-tuning alone reaches only 58\% (100
scenarios) or 34.76\% (164 scenarios).
This indicates that, for the current task scope and benchmark design, the retrieval and prompt-engineering pipeline is sufficient to bridge the gap between a general-purpose LLM and a reliable power-flow agent.
Fine-tuning does not degrade performance when combined with retrieval, but it does not provide measurable additional benefit either.

One interpretation is that the fine-tuning dataset, while curated and validated, is not yet large enough to cover the full range of ANDES API patterns.
Producing substantially more fine-tuning data, particularly multi-turn sequences with cumulative modifications, may be needed before fine-tuning can match retrieval on these tasks.
This observation is consistent with the finding in~\cite{jia2025enhancing} that retrieval and tool feedback are more impactful than model scale alone for code generation.

\subsubsection{Retrieval Granularity Matters}

It is noted that the effectiveness of retrieval is highly sensitive to the chunking strategy.
The initial implementation used sentence-level chunks, which fragmented the ANDES manual into isolated pieces and frequently broke the structural context needed for correct code generation.
Switching to the window-level passage scheme provides the 100\% pass above.
For API- and workflow-heavy domains like power system simulation, preserving local context in each retrieved passage is more important than optimizing the embedding model.

\subsubsection{Multi-Turn Continuity as a Differentiator}

The turn-level analysis reveals that Base~OpenAI and Fine-tuned both degrade across turns.
In contrast, retrieval-based modes maintain 100\% at every turn.
This is because the retrieval pipeline explicitly injects
i)~the active ANDES case identifier and continuity state,
ii)~the full case index inventory, and
iii)~the conversation compaction summary
into every prompt. This provides the LLM with the information it needs to generate code that preserves prior modifications.
Without this explicit context, the LLM must reconstruct the cumulative state from the conversation history alone, which becomes increasingly unreliable as the number of modifications grows.

\subsubsection{Self-Evolution without Retraining}

The 164-scenario before/after comparison demonstrates that the self-evolution mechanism can recover from a 60.98\% pass rate to 100.0\% without any model retraining.
The root cause of the pre-evolution failures was a mismatch between the ANDES line-outage API and the patterns embedded in the agent's prompt templates.
The failure-attribution step correctly identified this as the dominant signature, including \texttt{corridor\_outage\_language} and \texttt{line\_outage\_api\_guardrail}, and the resulting constraints updated the prompt to use the correct API form.

These results support a practical development workflow for power-system agents, which include six key steps:
\begin{itemize}
  \item Expand the benchmark with more complex tasks.
  \item Run the agent and collect failures.
  \item Use failure clustering to locate the dominant bottleneck.
  \item Determine whether the bottleneck is model capability, agent workflow, or verifier mismatch.
  \item Update the evolution profile accordingly.
  \item Rerun the same suite to measure recovery.
\end{itemize}
This cycle can be repeated with the same base LLM until the agent reaches the desired level of reliability and coverage.

% ========================================================================================
% Section VI: Conclusion
% ========================================================================================

\section{Conclusion}
\label{sec:conclusion}

This paper presents \emph{PFAgent}, a tractable and self-evolving power-flow agent for interactive grid analysis on ANDES.
By integrating intent parsing, knowledge retrieval, tool execution, structured reporting, self-evolution, and an AI-assisted fixing loop, the proposed framework enables text-to-simulation workflows.
The case studies show that retrieval-augmented grounding is the key factor for reliable multi-turn performance, while the self-evolution mechanism can effectively recover from dominant failure modes through prompt- and rule-level updates without retraining the base model.
The results suggest that reliable power-system agents should be built as execution-based and continuously improvable systems, rather than being treated as standalone language models.

The current study is limited to steady-state power flow analysis and the benchmark scope constructed in this work.
Future work will extend the framework to optimal power flow, dynamic and stability analysis, and cross-simulator tasks.

% ------------------------- Acknowledgement ------------------------
\section*{Acknowledgment}

This work was supported by the startup fund of the Department of Electrical and Computer Engineering, Kansas State University.
% The authors would like to acknowledge the assistance of ChatGPT-5.4 in polishing the language of this manuscript.
% The authors have reviewed and take full responsibility for the content of the text.

% ------------------------------------ Reference -----------------------------
\bibliographystyle{IEEEtran}
\bibliography{ref}

% that's all folks
\end{document}